\pdfoutput=1
\PassOptionsToPackage{obeyspaces}{url}

\documentclass[12pt]{iopart}

\expandafter\let\csname equation*\endcsname\relax
\expandafter\let\csname endequation*\endcsname\relax
\usepackage{amsmath,amssymb}

\usepackage{graphicx}

\usepackage{bm}
\usepackage{multirow}
\usepackage[normalem]{ulem} % \sout に必要
\usepackage{xcolor}
\usepackage{cite}

\usepackage[colorlinks=true,allcolors=blue]{hyperref}

\newcommand{\Eprint}[2]{\href{#1}{\urlstyle{same}\nolinkurl{#2}}}

\newcommand{\pd}{\partial}
\newcommand{\diag}{\qopname\relax o{diag}}
\newcommand{\sign}{\qopname\relax o{sgn}}

\newcommand{\reals}{\mathbb{R}}

\newcommand{\n}{\mathfrak{n}}

\newcommand{\calC}{{\cal C}}
\newcommand{\hatS}{\hat{S}}

\newcommand{\AHM}{\overline{\text{HM}}}

\def\({\left(}  
\def\){\right)} 
\def\[{\left[}
\def\]{\right]} 
\def\<{\left<} 
\def\>{\right>}

\newcommand{\mn}[1]{\textcolor{red}{#1}}

\begin{document}

\title[Hyperbolic monopoles in $SU(3)$ Yang-Mills theory]
{Hyperbolic monopole-type solutions in $SU(3)$ Yang-Mills theory}

\author{Yuki Amari}
\address{International Institute for Sustainability with Knotted Chiral Meta Matter (WPI-SKCM$^2$), Hiroshima University, 1-3-1 Kagamiyama, Higashi-Hiroshima, Hiroshima 739-8531, Japan}
\address{
Research and Education Center for Natural Sciences,
Keio University,
4-1-1 Hiyoshi, Kohoku, Yokohama,
Kanagawa 223-8521, Japan
}

% \ead{your-email@example.com}

\vspace{10pt}
\begin{indented}
\item[]\today
\end{indented}

\begin{abstract}
We construct $S^1$-invariant solutions of the $SU(3)$ pure Yang–Mills theory on four-dimensional Euclidean space using the Cho–Faddeev–Niemi decomposition and a harmonic map from $S^2$ into the flag manifold $F_3=SU(3)/U(1)^2$. The resulting ansatz reduces the full Yang–Mills equations to coupled radial equations. 
Solving the coupled equations, we derive two types of solutions: analytic self-dual solutions interpreted as noninteracting clusters of hyperbolic monopoles and non-self-dual numerical solutions that can be interpreted as hyperbolic monopole–antimonopole bound states. 
In the large-mass limit of the hyperbolic monopole–antimonopole bound states, their normalized action approaches the energy of the corresponding non-Bogomolny $SU(3)$ monopole in flat space.
\end{abstract}

\vspace{2pc}

%\submitto{J. Phys. A: Math. Theor.}
%%%%%%%%%%%%%%%%%%%%%%%%%%%%%%%%%%%%%%%%%%%%%%%%%%%%%%%%%%%%%

\section{Introduction}
\label{sec:introduction}

Classical solutions of Yang-Mills (YM) theory provide concrete realizations of the nonlinear and topological structures of non-Abelian gauge fields and describe nonperturbative degrees of freedom.
Among the most important examples are instantons, finite-action solutions in four-dimensional Euclidean space that satisfy the (anti-)self-duality equations and saturate the topological lower bound on the action.
The unit-charge Belavin-Polyakov-Schwartz-Tyupkin instanton~\cite{Belavin:1975fg} is spherically symmetric in four dimensions. Instantons invariant under an $SO(3)$ rotational symmetry about the $x^4$ axis are known as Witten instantons, or hyperbolic vortices, because their symmetry reduction yields the Bogomolny equations for vortices on the hyperbolic plane $\mathbb{H}^2$~\cite{Witten:1976ck}. Similarly, instantons possessing an $SO(2)$ symmetry correspond to monopoles on $\mathbb{H}^3$ and therefore such a solution is called hyperbolic monopole (HM)~\cite{atiyah1987magnetic,Chakrabarti:1984sm}.

In contrast to instantons, non-self-dual solutions of the full YM
equation remain comparatively poorly understood. Such solutions are saddle points of the YM action rather than absolute minima in a given topological sector. Their construction therefore requires solving the second-order YM equation without the integrable structure available in the self-dual sector.
In $SU(3)$ YM theory, symmetry reductions have produced non-self-dual
configurations described by coupled Abelian Higgs variables
\cite{Burzlaff:1980hr,Schiff:1991bu}, as well as finite-energy monopole
solutions that do not satisfy the Bogomolny equations
\cite{Burzlaff:1981hs,Ioannidou:1999xq}.

In this paper, we study $S^1$-invariant HM-type solutions in $SU(3)$ YM theory, especially non-self-dual solutions that are hyperbolic counterparts of monopole solutions studied in Ref.~\cite{Burzlaff:1981hs}. 
To obtain such solutions, we employ the Cho-Faddeev-Niemi (CFN) decomposition~\cite{Faddeev:1998eq,Faddeev:1998yz,Faddeev:1999cj,Shabanov:1999xy,Shabanov:1999uv,Kondo:2005eq} to construct an ansatz. The CFN decomposition is a parametrization of the gauge potential separating Abelian components along local
color directions and off-diagonal components tangent to the color orbit. It was first developed to describe low-energy effective degrees of freedom and to derive an effective model of the YM theory, called the Skyrme-Faddeev model \cite{Faddeev:1976pg,Faddeev:1996zj}.
However, it is known that the CFN decomposition contains Witten's axially symmetric ansatz \cite{Witten:1976ck} and therefore can reproduce axially symmetric multi-instantons as well as meron-antimeron \cite{deAlfaro:1976qet} and its generalization in Minkowski spacetime \cite{Luscher:1977cw,Schechter:1977qg}.
By separating variables using a harmonic map from $S^2\subset \mathbb{R}^4$ to the flag manifold $F_3=SU(3)/U(1)^2$ \cite{negreiros1988some,Bykov:2015pka,Amari:2017qnb}\footnote{This manifold is sometimes referred to as $F_2$ instead of $F_3$ \cite{Kondo:2008xa}.} and imposing $S^1$ symmetry, we reduce the full YM equation to a coupled system of ODEs for the radial profile functions.

By solving the reduced system, we derive two classes of solutions.  
The first is a family of analytic self-dual solutions describing clusters of HMs with zero binding energy, whereas the second is non-self-dual solutions describing bound states of hyperbolic monopole and antimonopole (HM-$\AHM$).
These interpretations follow from the magnetic weights of the solutions.
This paper is organized as follows.

Section~\ref{sec:formulation} defines the action and topological charge and
introduces the CFN decomposition of the $SU(3)$ gauge potential.
Section~\ref{sec:harmonic-map} presents the harmonic-map ansatz and derives the coupled radial ODEs from the YM equations.
In Sec.~\ref{sec:HmM}, we analytically construct HmM solutions describing noninteracting clusters of HMs.
Section~\ref{sec:HM-AHM} presents numerical non-self-dual solutions
interpreted as hyperbolic monopole-antimonopole bound states.
Finally, Sec.~\ref{sec:summary} summarizes our results.

\section{Cho-Faddeev-Niemi decomposition of the \texorpdfstring{$SU(3)$}{SU(3)} gauge potentials}
\label{sec:formulation}

In this paper, we consider the $SU(3)$ pure YM theory on $\mathbb{R}^{4}$. This section defines the action and introduces the parametrization of the gauge potential called the CFN decomposition.

Employing the $SU(3)$ generators $T_m=\lambda_m/2$ with the Gell-Mann matrices  $\lambda_m~(m=1,2,...,8)$, the action density of the theory in Euclidean signature is given by
\begin{equation}
{\cal S}=\frac{1}{2} \operatorname{Tr}\left(F_{\mu \nu} F_{\mu \nu}\right) \ ,
\label{eq:YMaction}
\end{equation}
with $SU(3)$ gauge potential $A_\mu = A^m_\mu T_m$ and the field strength 
\begin{equation}
F_{\mu \nu}=\partial_{\mu} A_{\nu}-\partial_{\nu} A_{\mu}-i\left[A_{\mu}, A_{\nu}\right]  \ .
\end{equation}
The topological charge is defined as
\begin{equation}
Q=\frac{1}{16 \pi^{2}} \int d^{4} x \operatorname{Tr}\left(F_{\mu \nu} \tilde{F}_{\mu \nu}\right) 
\end{equation}
with
\begin{equation}
\tilde{F}_{\mu \nu}=\frac{1}{2} \varepsilon_{\mu \nu \alpha \beta} F_{\alpha \beta} \ .
\end{equation}
For a finite-action configuration, the gauge potential approaches a pure gauge at spatial infinity.  It therefore defines a map $S^{3}_{\infty}\to SU(3)$, and the topological charge is an integer labeled by homotopy class
$ \pi_{3}\left(SU(3)\right) = \mathbb{Z}$ .

The Euler-Lagrange equation of the action \eqref{eq:YMaction}, called the YM equation, is given by
\begin{equation}
\partial_{\mu} F_{\mu \nu}-i\left[A_{\mu}, F_{\mu \nu}\right]=0 \ .
\label{eq:YMeq}
\end{equation}
In this paper, we consider the full YM equation \eqref{eq:YMeq} rather than the (anti)self-dual equation $F_{\mu\nu}=\pm\tilde{F}_{\mu\nu}$, because we consider not only minimum action solutions in a given topological sector but also saddle point solutions of the action \eqref{eq:YMaction}. 
Since we cannot apply powerful techniques like the Atiyah-Drinfeld-Hitchin-Manin construction \cite{Atiyah:1978ri} to solve the full YM equation, we construct an educated ansatz and reduce the full YM equation to a coupled ODE system.

To construct an educated ansatz, we employ the following parametrization of the gauge potential called the CFN decomposition:
\begin{equation}
    A_{\mu}=C_{\mu}^{a} \n_{a} +i\left[\n_{a}, \partial_{\mu} \n_{a}\right] + i \rho_{a b}\left[\n_{a}, \partial_{\mu} \n_{b}\right]+\sigma_{a b}\left\{\n_{a}, \partial_{\mu} \n_{b}\right\}
    \label{eq:SU(3)CFN}
\end{equation}
where $a, b=1,2$. 
The color(-direction) fields $\n_a$ can be defined by 
\begin{equation}
\n_{a}=U h_{a} U^{\dagger} 
\end{equation}
with $U \in SU(3)$ and 
\begin{equation}
    h_1=\frac{\lambda_3}{2}, \qquad
    h_2=\frac{\lambda_8}{2}.
\end{equation}
The matrices $h_a$ form the Cartan-Weyl basis of the $su(3)$ algebra together with 
\begin{equation}
e_{ \pm 1}=\frac{1}{2}\left(\lambda_{1} \pm i \lambda_{2}\right), \quad
e_{ \pm 2}=\frac{1}{2}\left(\lambda_4 \mp i \lambda_{5}\right), \quad
e_{ \pm 3}=\frac{1}{2}\left(\lambda_{6}\pm i \lambda_{7}\right) \ .
\end{equation}
Moreover, $C_{\mu}^{a}$ are Abelian gauge potentials, $\rho_{a b}$ and $\sigma_{a b}$ are real scalar fields.
%The properties of the Cartan-Weyl basis are collected in Appendix.~\ref{app:Cartan-Weyl}.

The fields $\rho_{ab}$ and $\sigma_{ab}$ contain eight real scalar components. However, only six components are independent because of the two identities  
\begin{align}
    &\left[\n_{1}, \partial_{\mu} \n_{2}\right]-\left[\n_{2}, \partial_{\mu} \n_{1}\right] =0 \ ,
    \\
    & \left\{\n_{1}, \partial_{\mu} \n_{1}\right\}+\left\{\n_{2}, \partial_{\mu} \n_{2}\right\}=0 \ .
\end{align}
These six components can be organized into the three complex fields $\phi_{k} ~ (k=1,2,3)$. 
We perform the gauge transformation
\begin{align}
 A_{\mu} \rightarrow A^U_{\mu}=U^{\dagger} A_{\mu} U+i U^{\dagger} \partial_{\mu} U  \ .
\end{align}
Then, one can write the gauge potential as
\begin{align}
A^U_{\mu}=  \sum_{a}\left(C_{\mu}^{a}+\omega_{\mu}^{a}\right) h_{a}+i\sum_{k}\left( \phi_{k} \eta_{\mu}^{k} e_{k}- \bar{\phi}_{k} \bar{\eta}_{\mu}^{k} e_{-k}\right)  \ ,
\label{eq:gauge_pot_gtransf}
\end{align}
where $\bar{\phi}_k$ stands for the complex conjugate of $\phi_k$. Here, we decomposed the pure gauge part in terms of the Cartan-Weyl basis as
\begin{equation}
i U^{\dagger} \partial_{\mu} U=\sum_{a=1}^2 \omega_{\mu}^{a} h_{a}+\sum_{k=1}^3\left(\eta_{\mu}^{k} e_{k}+\bar{\eta}_{\mu}^{k} e_{-k}\right) \ .
\end{equation}
The complex scalar fields $\phi_{k}$ in Eq.~\eqref{eq:gauge_pot_gtransf} are given by
\begin{equation}
\phi_{k}=\sum_{a, b}\left(\sigma_{a b} \alpha_{b}^{k} \beta_{a}^{k}+i \rho_{a b} \alpha_{a}^{k} \alpha_{b}^{k}\right) \ , 
\end{equation}
where $\alpha_a^k$ and $\beta_a^k$ respectively stand for the $a$-th component of the root vector $\alpha^k$ and the anticommutator coefficient vector $\beta^k$, defined through
\begin{equation}
\left[h_{a}, e_{k}\right]=\alpha_{a}^{k} e_{k} ,
\qquad
\left\{h_{a}, e_{k}\right\}=\beta_{a}^{k} e_{k} \ ,
\end{equation}
for each $k=\pm 1, \pm 2, \pm 3$. 
The $U(1)^2$ gauge transformation of $A^U_\mu$ by $g=e^{i\gamma_a h_a}$ is equivalent to the transformation
\begin{equation}
    \phi_k\to\phi_ke^{i\vartheta_k},
    \qquad
    {\cal C}_{\mu}^{k} \equiv \alpha_{a}^{k} C_{\mu}^{a}\to {\cal C}_{\mu}^{k} + \partial_\mu \vartheta_k \ ,
\end{equation} 
where $\vartheta_k=\gamma_a\alpha_a^k$ satisfying $\vartheta_1+\vartheta_2+\vartheta_3=0$. Due to this transformation property, the pair $\{\phi_k, {\cal C}^k_\mu\}$ can be regarded as an Abelian-Higgs multiplet subject to $\sum_k \vartheta_k=0$.
%The explicit expressions for the  field strength, the action and topological charge densities, and the full YM equations in terms of the CFN variables are given in \ref{app:CFN-details}. 

The equation derived by variations with respect to the variables in Eq.~\eqref{eq:SU(3)CFN} is called the Faddeev-Niemi (FN) equation. Note that the FN equation is not equivalent to the YM equation \cite{Evslin:2010sb,Niemi:2010ms}. We would like to stress that in this paper, we solve the YM equation rather than the FN equation, and the CFN decomposition is just used to construct an ansatz.

\section{Construction of ansatz}
\label{sec:harmonic-map}

In the $SU(2)$ case, it is known that the CFN connection can reproduce several classical solutions in the pure YM theory such as Wu-Yang monopole, Witten instanton and Hyperbolic monopole~\cite{Cho:1979nv,Cho:1980nx,Faddeev:1999pr,Kondo:2025dtx,kondo2026quarkconfinementunifiedmagnetic}, when the color field is a harmonic map from $S^2\subset \mathbb{R}^4$ to $S^2$. As a generalization, in this section, we introduce the harmonic map ansatz for the $SU(3)$ case, and write down the resulting equations of motion, action and topological charge.

We first specify the coordinate system to define the base space $S^2\subset \mathbb{R}^4$. 
To transparently describe hyperbolic monopole-type solutions, we introduce the ring coordinates $(\xi, \psi, \theta, \chi)$ through
\begin{equation} 
x_1+ix_2 = \frac{R\sinh\xi\sin\theta}{\cosh\xi+\sinh\xi\cos\theta}e^{i\chi} 
\ , \qquad
x_3+ix_4 = \frac{R}{\cosh\xi+\sinh\xi\cos\theta}e^{i\psi} \ ,
\end{equation}
where $R$ is a constant, $\xi \in[0, \infty), \theta\in[0,\pi]$, and $\chi, \psi \in[0,2 \pi)$. 
Introducing the complex coordinate $z=\tan(\theta/2)e^{i\chi}$, one can write the line element as
\begin{align}
    ds^{2}&=dx_1^2+dx_2^2+dx_3^2+dx_4^2 =\lambda^{2}\left[d \psi^{2}+d \xi^{2}+4 \sinh ^{2} \xi \frac{d z d \bar{z}}{\left(1+|z|^{2}\right)^{2}}\right]
\end{align}
with 
\begin{equation}
    \lambda=\frac{R\left(1+|z|^{2}\right)}{\cosh \xi\left(1+|z|^{2}\right)+\sinh \xi\left(1-|z|^{2}\right)} \ .
\end{equation}
Here, $\bar{z}$ stands for the complex conjugate of $z$.
This metric is conformally equivalent to that of $S^1\times \mathbb{H}^3$, where $\mathbb{H}^3$ is the three-dimensional hyperbolic space, and the $\mathbb{H}^3$ part is described by a ball model.

In the $SU(3)$ case, the gauge potential \eqref{eq:SU(3)CFN} contains the two color fields $\n_1$ and $\n_2$. These take values on the flag manifold $F_3=SU(3)/U(1)^2$. 
We take the pair of color fields to define a harmonic map from $S^2$ parametrized by $\{z,\bar{z}\}$ to $F_3$. 
We have two types of harmonic maps: trivial embedding type and full map type.
For the trivial embedding, an $SU(2)$ matrix is embedded as a $2\times2$ block of $U\in SU(3)$.
For example,
\begin{align}
    & U=\frac{1}{\sqrt{1+|z|^{2}}}\left(\begin{array}{ccc}
1 & 0 & -\bar{z} \\
0 & \sqrt{1+|z|^{2}} & 0 \\
z & 0 & 1
\end{array}\right) \ ,
\label{eq:harmonic-map_embedding}
\end{align}
where the second column is the constant vector, independent of $z$ and $\bar{z}$.
Eq.~\eqref{eq:harmonic-map_embedding} represents a holomorphic map into a $CP^1$ subspace of $F_3$, and yields the $SU(2)$ HM solution as a trivial embedded solution.
%We describe the construction of an embedding type HM solution with Eq.~\eqref{eq:harmonic-map_embedding}  in the Appendix \ref{sec:embedding} as a reference.

Our main focus in this paper is the full map type.
Such a configuration can be given by the following $SU(3)$ matrix:
\begin{align}
    & U=\frac{1}{1+|z|^{2}}\left(\begin{array}{ccc}
1 & -\sqrt{2} \bar{z} & -\bar{z}^{2} \\
\sqrt{2} z & 1-|z|^{2} & \sqrt{2} \bar{z} \\
-z^{2} & -\sqrt{2} z & 1
\end{array}\right) \ .
\label{eq:harmonic-map_genuine}
\end{align}
Note that this type of the harmonic map ansatz is equivalent to those used in Refs.~\cite{Ioannidou:1999iw,Ioannidou:1999xq}, although the authors refer to it as a harmonic map from $S^2$ to $CP^2$ rather than $F_3$.
As we shall see, Eq.~\eqref{eq:harmonic-map_genuine} can describe HM-$\overline{\text{HM}}$ configurations\footnote{In the cylindrical coordinate system defined by the line element
\begin{equation}
    ds^{2}=d \tau^{2}+d r^{2}+4 r^2  \frac{d z d \bar{z}}{\left(1+|z|^{2}\right)^{2}} \ ,
\end{equation}
with $\tau\in(-\infty,\infty)$ and $r\in[0,\infty)$, one can construct an $SU(3)$ generalization of Witten instanton \cite{Witten:1976ck} that can be interpreted as a bound state of instanton -- anti-instanton \cite{Burzlaff:1980hr,Schiff:1991bu} using Eq.~\eqref{eq:harmonic-map_genuine}.}, which is the hyperbolic counterpart of non-BPS $SU(3)$ monopoles studied in Ref.~\cite{Burzlaff:1981hs}.

Similar to the standard HM case, we further assume $S^1$-invariance along the angle $\psi$ and adopt an ansatz of the form 
\begin{equation}
    C_{\psi}^{a}=C_{\psi}^{a}(\xi), \quad
    C_{\xi}^{a}=C_{z}^{a}=C_{\bar{z}}^{a}=0, \quad
    \phi_{k}=\bar{\phi}_{k}=\varphi_{k}(\xi) \ .
    \label{eq:S1-inv_ansatz}
\end{equation}
With these assumptions, the YM equation \eqref{eq:YMeq} reduces to
\begin{align}
& 0=\partial_{\xi}\left[\sinh ^{2} \xi ~ \partial_{\xi} C_{\psi}^{1}\right]-4 C_{\psi}^{1}\varphi_{1}^{2}-\left(C_{\psi}^{1}-\sqrt{3} C_{\psi}^{2}\right)\varphi_{3}^{2}  \ ,
\label{eq:eom_gen_C1}
\\
& 0=\partial_{\xi}\left[\sinh ^{2} \xi ~ \partial_{\xi} C_{\psi}^{2}\right]+\sqrt{3}\left(C_{\psi}^{1}-\sqrt{3} C_{\psi}^{2}\right)\varphi_{3}^{2}  \ ,
\label{eq:eom_gen_C2}
\\
& 0=\partial_{\xi}^{2} \varphi_{1}-\left(C_{\psi}^{1}\right)^{2} \varphi_{1}+\frac{\varphi_{1}}{\sinh ^{2} \xi}\left(1-2\varphi_{1}^{2}+\varphi_{3}^{2}\right) \ ,
\label{eq:eom_gen_phi1}
\\
& 0=\partial_{\xi}^{2} \varphi_{3}-\frac{1}{4}\left(C_{\psi}^{1}-\sqrt{3} C_{\psi}^{2}\right)^{2} \varphi_{3}+\frac{\varphi_{3}}{\sinh ^{2} \xi}\left(1+\varphi_{1}^{2}-2\varphi_{3}^{2}\right) \ .
\label{eq:eom_gen_phi3}
\end{align}
In addition, the action and topological charge can be cast into the form
\begin{align}
   & \begin{aligned}
S=16 \pi^2 \int  d \xi & {\left[\frac{1}{4} \sinh ^{2} \xi \sum_{a}\left(\partial_{\xi} C_{\psi}^{a}\right)^{2}+\left(\partial_{\xi} \varphi_{1}\right)^{2}+\left(\partial_{\xi} \varphi_{3}\right)^{2}\right.} \\
& \qquad
+\left(C_{\psi}^{1}\right)^{2}\varphi_{1}^{2}
+\frac{1}{4}\left(C_{\psi}^{1}-\sqrt{3} C_{\psi}^{2}\right)^{2}\varphi_{3}^{2}  
\\
&\qquad \left.+\frac{1}{\sinh^2  \xi} \left(1-\varphi_{1}^{2}-\varphi_{3}^{2}-\varphi_{1}^{2} \varphi_{3}^{2}+\varphi_{1}^{4}+\varphi_{3}^{4}\right)\right]
\label{eq:reduced_action_gen}
\end{aligned}
\\
&
Q  =-2 \int d \xi \sum_{a=1}^2\sum_{k=1,3}\left[\left(1-\varphi_{k}^{2}\right) \alpha_{a}^{k} \partial_{\xi} C_{\psi}^{a}-2 \alpha_{a}^{k} C_{\psi}^{a} \varphi_{k} \partial_{\xi} \varphi_{k}\right] 
\end{align}
where we used $\displaystyle \int \frac{dzd\bar{z}}{(1+|z|^2)^2}=2\pi i$.
Eqs.~\eqref{eq:eom_gen_C1}-\eqref{eq:eom_gen_phi3} can be derived by variations of the action \eqref{eq:reduced_action_gen} with respect to the fields $\{C_\mu^a, \phi_k\}$. However, we would like to stress that these equations ~\eqref{eq:eom_gen_C1}-\eqref{eq:eom_gen_phi3} are obtained from the full YM equation \eqref{eq:YMeq} rather than the FN equation. 

In this paper, we consider, for simplicity, the case in which the four radial functions admit a consistent reduction to two independent functions.
We suppose that 
\begin{align}
    \varphi_1= f,\quad 
    \varphi_3= u f, 
    \qquad 
    C_{\psi}^{1}= \calC, \quad
    C_{\psi}^{2}=v \calC \ . 
    \label{eq:2function_reduction}
\end{align}
where $\{u,v\}$ are constants, and $\{f,\calC\}$ are real functions\footnote{The four-equation system also reduces to a two-equation system when 
$\varphi_1=0, C_\psi^1=-C_\psi^2/\sqrt{3}$.
In this case, the action density diverges at $\xi=0$, because the potential term is proportional to $V=1-\varphi_3^2+\varphi_3^4=(\frac{1}{2}-\varphi_3^2)^2+ \frac{3}{4}>0$ and $V/\sinh^2\xi\to\infty$ as $\xi\to0$. Thus, we do not consider such a case in this paper.
}. 
Substituting Eq.~\eqref{eq:2function_reduction} into Eqs.~\eqref{eq:eom_gen_phi1} and \eqref{eq:eom_gen_phi3}, one obtains 
\begin{align}
& 0=\partial_\xi^2f-\calC^2 f + \frac{f}{\sinh^2\xi}(1-2f^2+u^2f^2) \ , \\
& 0= \partial_\xi^2f-\frac{1}{4}(1-\sqrt{3}v)^2\calC^2 f + \frac{f}{\sinh^2\xi}(1+f^2-2u^2f^2) \ .
\end{align}
Comparing the coefficients in the equations, one finds that these two equations are identical if 
\begin{equation}
    (u,v) = (\pm 1,~\sqrt{3}) \ ,
    \label{eq:cond_HmM}
\end{equation}
or
\begin{equation}
    (u, v) = \left(\pm 1,~-\frac{1}{\sqrt{3}}\right) \ .
    \label{eq:cond_HM-HAM}
\end{equation}
For both cases, Eqs.~\eqref{eq:eom_gen_C1} and \eqref{eq:eom_gen_C2} are consistent.
Note that we can restrict ourselves to $u=1$ without loss of generality because the sign of $u$ can be flipped by a global gauge transformation.
In the following sections, we solve equations for $\{f,\calC\}$ for both cases \eqref{eq:cond_HmM} and \eqref{eq:cond_HM-HAM}.

\section{Hyperbolic multi-monopoles}
\label{sec:HmM}

Here, we construct HM-type solution for the case \eqref{eq:cond_HmM}, which means 
\begin{align}
    \varphi_1=\varphi_3\equiv f, \qquad 
    C_{\psi}^{1}=\frac{C_{\psi}^{2}}{\sqrt{3}}\equiv \calC \ .
    \label{eq:ansatz_HmM}
\end{align}
We show that in this case, we can construct such solutions using the BPS trick and the solutions describe hyperbolic multi-monopoles (HmM). 
This class of solutions is equivalent to the hyperbolic monopoles obtained in Ref.~\cite{Ioannidou:1999iw}.
Although these solutions are not new, we discuss their properties in detail to facilitate comparison with the solutions constructed in the next section.

\subsection{Formulation}

We first write down the equations of motion, action, and the topological charge for the ansatz \eqref{eq:ansatz_HmM}. 
For the ansatz, Eqs.~\eqref{eq:eom_gen_C1} and \eqref{eq:eom_gen_C2} become the same equation of the form
\begin{equation}
    \partial_{\xi}\left[\sinh ^{2}\xi ~ \partial_{\xi} \calC \right]-2 \calC f^{2}=0 \ ,
    \label{eq:eom_HmM_Cf}
\end{equation}
and Eqs.~\eqref{eq:eom_gen_phi1} and \eqref{eq:eom_gen_phi3} become
\begin{equation}
    \partial_{\xi}^{2} f -\calC^{2} f+\frac{f\left(1-f^{2}\right)}{\sinh ^{2} \xi}=0 \ . 
    \label{eq:eom_HmM_fC}
\end{equation}
The action and topological charge can be written as
\begin{align}
&S=32 \pi^{2} \int d \xi \left[  \frac{1}{2} \sinh^{2} \xi\left(\partial_\xi \calC \right)^{2} +(\partial_\xi f)^{2}+\calC^{2} f^{2}+\frac{1}{2} \frac{\left(1-f^{2}\right)^{2}}{\sinh^{2} \xi}\right] \ ,
\label{eq:action_HmM}
\\
& Q=-4\int d \xi ~ \partial_{\xi}\left[\left(1-f^{2}\right) \calC \right] \ ,
\label{eq:charge_HmM}
\end{align}
Note that the equations \eqref{eq:eom_HmM_Cf} and \eqref{eq:eom_HmM_fC} can be obtained as the EL equations of the action \eqref{eq:action_HmM}. 

The action and topological charge are exactly four times larger than those of the trivial embedding configuration, %presented in \ref{sec:embedding},
 although the fields $\{f,\calC\}$ have different interpretations.
This factor of four arises because the ansatz \eqref{eq:ansatz_HmM} realizes the principal embedding of $SU(2)$ into $SU(3)$.
For the ansatz, the gauge potentials are given as
\begin{align}
& A^U_{\psi}=\calC \left(h_{1}+\sqrt{3} h_{2}\right) \ , \notag\\
& A^U_{\xi}=0 \ , 
\label{eq:gaugepot_HmM}\\
& A^U_{z}=\frac{i \bar{z}}{1+|z|^{2}}\left(h_{1}+\sqrt{3} h_{2}\right)-\frac{\sqrt{2}f}{1+|z|^{2}} \left(e_{-1}-e_{-3}\right) \notag \ .
\end{align}
One can see that the gauge potential takes values in the three-dimensional irreducible representation of $su(2)$ in $su(3)$, the principal $su(2)$ subalgebra, spanned by the generators
\begin{align}
J_3&=h_1+\sqrt{3}h_2 \ , 
\qquad
J_\pm
=\sqrt{2}\left(e_{\pm1}-e_{\pm3}\right).
\label{eq:generator_principal_SU(2)}
\end{align}
On the other hand, the generators of the trivially embedded $su(2)$ subalgebra may be chosen as
\begin{equation}
I_3=\frac{1}{2}\left(h_1+\sqrt{3}h_2\right) \, \qquad I_\pm=e_{\mp2}.
\end{equation}
The factor of four originates from the different normalizations of the two embeddings:
\begin{equation}
\Tr\left(J_3^2\right)
=4\Tr\left(I_3^2\right) \ , 
\qquad
\Tr\left(J_+ J_- \right)
=4\Tr\left(I_+ I_-\right) \ .
\end{equation}

\subsection{Construction of the solutions}

The reduced action differs from that of the standard HM only by an overall factor and yields the same EL equations. With the same boundary conditions, the corresponding radial profile functions also coincide. 
%We follow the construction of the standard HM solutions \cite{Chakrabarti:1984sm}.
Since Eqs.~\eqref{eq:eom_HmM_Cf} and \eqref{eq:eom_HmM_fC} are the EL equations associated with the action \eqref{eq:action_HmM}, any configuration satisfying the resulting first-order BPS equations automatically solves these second-order equations.
The BPS bound of the action is given by
\begin{align}
S= & 32 \pi^{2} \int d \xi\left[\frac{1}{2}\left(\sinh \xi \partial_{\xi} \calC \pm \frac{1-f^{2}}{\sinh \xi}\right)^{2}+\left(\partial_{\xi} f \mp \calC f\right)^{2}\right] \notag\\
&\qquad\mp 32 \pi^{2} \int d \xi\left[\left(1-f^{2}\right) \partial_{\xi} \calC -2 \calC f \partial_{\xi} f\right] \notag\\
\geq & 8 \pi^{2}|Q| 
\end{align}
Therefore, the BPS equations are
\begin{align}
 \partial_{\xi} \calC=\mp \frac{1-f^{2}}{\sinh ^{2} \xi}  
,\qquad
\partial_{\xi} f= \pm \calC f \ .
\end{align}
These equations are equivalent to the BPS equations for the HMs in the $SU(2)$ Yang-Mills theory and therefore we solve them following Ref.~\cite{Chakrabarti:1984sm}.
These imply
\begin{align}
& \calC = \pm \partial_{\xi} \log f , \\
& \partial_{\xi}^{2} \log f=-\frac{1-f^{2}}{\sinh ^{2} \xi} \ .
\label{eq:BPS_HmM_f}
\end{align}
Let us write 
\begin{equation}
    f=\frac{\sinh \xi}{\eta(\xi)}
\end{equation}
and substitute it into Eq.~\eqref{eq:BPS_HmM_f}. Then one gets
\begin{equation}
    \left(\partial_{\xi} \eta\right)^{2}-\eta \partial_{\xi}^{2} \eta=1 \ .
\end{equation}
This is solved by
\begin{equation}
    \eta=\frac{\sinh (\alpha \xi)}{\alpha}
\end{equation}
where $\alpha$ is a constant that we can restrict to be positive without loss of generality due to the invariance of $\eta$ under $\alpha\to-\alpha$. Summarizing the above discussion, we find that the solutions are given by
\begin{align}
& f=\frac{\alpha \sinh \xi}{\sinh (\alpha \xi)},  \label{eq:HmM_sol_f}\\
& \calC= \pm(\operatorname{coth} \xi-\alpha \operatorname{coth}(\alpha \xi))
\label{eq:HmM_sol_calC}
\end{align}
When $\alpha<1$, the function $f$ diverges at large $\xi$ and therefore the action also diverges. If $\alpha=1$, the solution is a pure gauge solution. Finite action solutions are given by $\alpha>1$, and the action is
\begin{equation}
   S=32 \pi^{2}(\alpha-1) 
\end{equation}
In general, the parameter $\alpha$ can be continuous, and so is the action. However, the gauge potentials are finite and single-valued everywhere in $\mathbb{R}^4$ if and only if $\alpha$ is an integer, as we shall see later.

%%%%%%%%%%%%%%%%%%%%%%%%%%%%%%%%%%%%%%%%%%
\begin{figure}[t]
    \centering
    \includegraphics[width=1.0\linewidth]{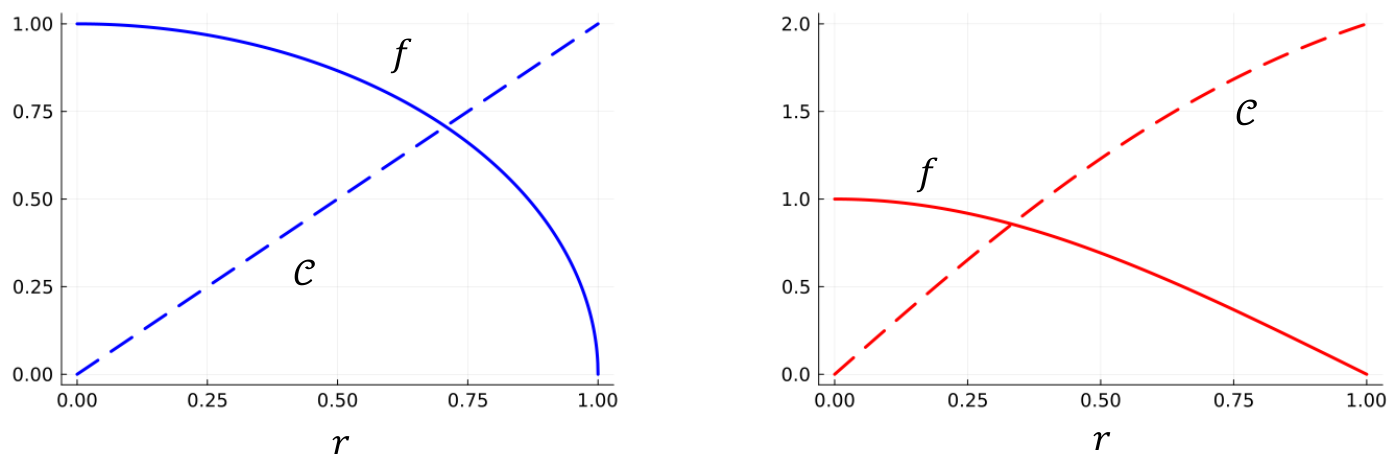}
    \caption{
        Radial profiles of the analytic solutions \eqref{eq:HmM_sol_f} and \eqref{eq:HmM_sol_calC} as functions of $r=\tanh\xi$.
        Solid and dashed curves represent $f$ and $\calC$ respectively. The left panel shows the solutions with $\alpha = 2$, and the right panel does those with $\alpha = 3$. The lower-sign branch of Eq.~\eqref{eq:HmM_sol_calC} is shown.}
    \label{fig:HmM_profile}
\end{figure}

\begin{figure}[t]
    \centering
    \includegraphics[width=0.5\linewidth]{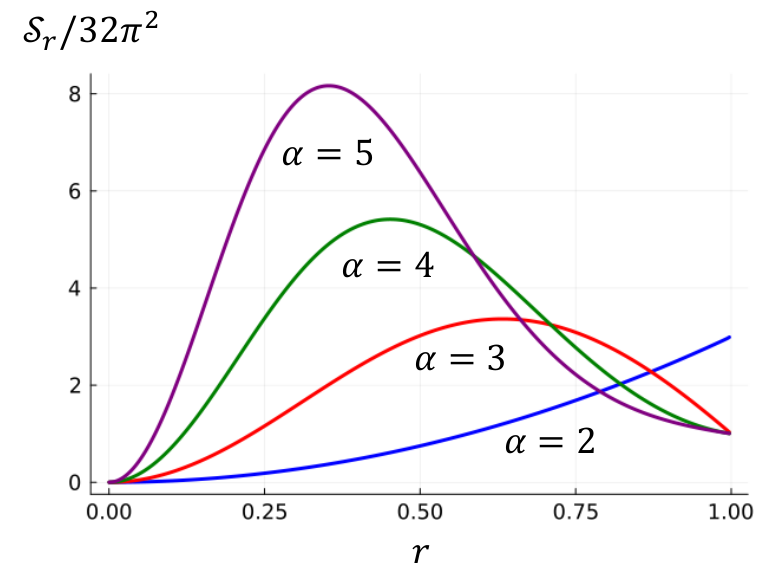}
    \caption{Normalized radial action density ${\cal S}_r/32\pi^2$ of the solutions with $\alpha=2,3,4,5$, where $S=\int dr {\cal S}_r$. For $\alpha=2$, the action density attains its maximum at $r=1$, corresponding to the limit $\xi\to\infty$. As $\alpha$ increases, the peak shifts toward smaller values of $r$.}
    \label{fig:HmM_action}
\end{figure}

%%%%%%%%%%%%%%%%%%%%%%%%%%%%%%%%%%%%%%%%%%

We show the profile of the solutions \eqref{eq:HmM_sol_f} and \eqref{eq:HmM_sol_calC} in Fig.~\ref{fig:HmM_profile} and the action distribution in Fig.~\ref{fig:HmM_action}. In these figures, we used the coordinate 
\begin{equation}
    r=\tanh \xi 
    \label{eq:normalized_coordinate}
\end{equation}
to facilitate comparison between these solutions and the HM-$\overline{\text{HM}}$ solutions presented in the next section.

The action density ${\cal S}_r$ is defined through $S=\int dr {\cal S}_r$. 
One can analytically find that in the limit $r\to 1$, ${\cal S}_r/32\pi^2\to 3$ for $\alpha=2$ and ${\cal S}_r/32\pi^2\to 1$ for $\alpha\geq 3$.

When $\alpha=n+1\in \mathbb{Z}$, the topological charge is given by $|Q|=4n$ and therefore $S=32\pi^2 n$, which is exactly $4n$ times larger than that of the single HM in the $SU(2)$ YM theory. Thus, this solution can be interpreted as a cluster of four HMs sitting on top of each other without interaction between them. This type of configuration tends to appear in $SU(3)$ symmetric field theory:
The $SU(3)$ Skyrme model ($F_3$ Skyrme-Faddeev model) possesses solutions interpreted as a cluster of four Skyrmions in the $SU(2)$ Skyrme model (Hopfions in the $CP^1$ Skyrme-Faddeev model) sitting on top of each other without interaction between them \cite{Ioannidou:1999mk,Amari:2018gbq}.

\subsection{Magnetic charge and magnetic weight}

We now evaluate the physical quantities characterizing these solutions in the three-dimensional theory obtained by dimensional reduction along the circle direction.
First, we define the quantities in a general form.
In the reduced model, $A_\psi^U$ can be viewed as an adjoint Higgs field and we define 
\begin{equation}
\Phi_\infty
\equiv\lim_{\xi\to\infty}A^U_\psi \ .
\label{eq:asymptotic_Higgs_HmM}
\end{equation}
By a Weyl transformation, we order the eigenvalues of the asymptotic Higgs
field $\Phi_\infty$ such that
\begin{equation}
    \Phi_\infty\to \widetilde{\Phi}_\infty
    =\diag(\mu_1,\mu_2,\mu_3),
    \qquad
    \mu_1\geq\mu_2\geq\mu_3 \ .
\end{equation}
 We introduce the matrix representations of the coroot 
\begin{equation}
    H_1=\diag(1,-1,0),
    \qquad
    H_2=\diag(0,1,-1),
    \qquad
    H_3=\diag(1,0,-1) \ .
\end{equation}
Here $H_1$ and $H_2$ are the simple coroots, while $H_3$ is the coroot
associated with their sum and satisfies $H_3=H_1+H_2$. 
Then, the mass of type-$j$ monopole is defined by
\begin{equation}
    m_j=\Tr(H_j\widetilde{\Phi}_\infty) \geq 0
\end{equation}
In addition, we define the magnetic charge matrix as
\begin{equation}
    Q_{\rm M}=\frac{1}{2\pi}\int_{S^2_\infty}
    F^U_{z\bar{z}}~dzd\bar{z}  \ .
\label{eq:charge_matrix_HmM}
\end{equation}
The same permutation must be applied simultaneously to the magnetic charge
matrix, $Q_{\rm M}\to\widetilde{Q}_{\rm M}$, and we expand in terms of $H_1$ and $H_2$ as
\begin{equation}
    \widetilde{Q}_{\rm M} = k_1 H_1+k_2 H_2 + k_3 H_3\ .
    \label{eq:charge_matrix}
\end{equation} 
Because $H_3=H_1+H_2$, the three coefficients $(k_1,k_2,k_3)$ are not uniquely determined. 
We therefore use the simple coroots $H_1$ and $H_2$ as
an independent basis and set $k_3=0$. In this convention, monopoles of type-1 and type-2 are the fundamental monopoles, whereas a type-3 monopole is a composite of one monopole of each fundamental type. 
Then, $(k_1,k_2)$ is called magnetic weight and can be interpreted as monopole number \cite{Goddard:1976qe,Weinberg:2012pjx}; a positive $k_j$ counts type-$j$ monopoles, while a negative $k_j$ counts the corresponding antimonopoles. The energy contribution of
the type-$j$ (anti)monopole is given by $8\pi^2|k_j|m_j$.

For the present solution, we obtain
\begin{align}
    &\Phi_\infty = \diag(n,0,-n) \ , \\
    &\widetilde{Q}_{\rm M}=\diag(2,0,-2) \ .
\end{align}
For $n>0$, the eigenvalues of $\Phi_\infty$ are already ordered, whereas for
$n<0$ the first and third entries of both matrices must be exchanged by the
same Weyl transformation. Thus, one obtains
\begin{align}
    &(m_1,m_2,m_3) = (|n|,|n|,2|n|) \ , \\
    &\widetilde{Q}_{\rm M}
    =2\sign(n)H_3=2\sign(n)(H_1+H_2) \ .
\end{align}
For $n>0$, the configuration possesses the magnetic weight $(k_1,k_2)=(2,2)$, and the two types of monopole have the same mass parameter $m_1=m_2=|n|$.
This indicates that the configuration is composed of two type-1 and two type-2 monopoles, for a total of four fundamental monopoles of mass $|n|$.
For $n<0$, the signs of the magnetic weights are reversed, and the corresponding constituents are antimonopoles.

\subsection{Finiteness and single-valuedness of the gauge potential}
\label{subsec:single-valuedness_HmM}

Let us show that the gauge potentials are finite and single-valued everywhere in $\mathbb{R}^4$ if and only if $\alpha$ is an integer, generalizing the discussion of the standard HM case \cite{Chakrabarti:1984sm}. 
Since both $\calC$ and $f$ are finite everywhere in $\mathbb{R}^4$, the gauge potentials \eqref{eq:gaugepot_HmM} also look finite everywhere at a glance. 
However, since the metric is singular at $\xi=\infty$, we need careful analysis in a coordinate system that is non-singular at $\xi=\infty$.
For this purpose, we introduce such coordinates $(\tau, \rho)$ through
\begin{equation}
\tau+i \rho=\tan \left(\frac{\psi+i \xi}{2}\right) \ .
\label{eq:tau_rho}
\end{equation}
Note that the limit $\xi \rightarrow \infty$ corresponds to $\tau=0, \rho=1$.
The metric in this coordinate system becomes conformally equivalent to the metric in cylindrical coordinates
\begin{equation}
d s^{2} \sim d \tau^{2}+d \rho^{2}+4 \rho^{2} \frac{d z d \bar{z}}{\left(1+|z|^{2}\right)^{2}} \ .
\end{equation}
Thus, the coordinate system can be regarded as a non-singular one, although the conformal factor diverges at $\tau=0, \rho=1$.

The $\tau$ and $\rho$ components of the gauge potentials are given as follows respectively:
\begin{align}
A^U_{\tau} & =\frac{\partial \psi}{\partial \tau} A^U_{\psi} =\frac{1}{4}\left[4+e^{-\xi} e^{i \psi}+e^{\xi} e^{-i \psi}+e^{\xi} e^{i \psi}+e^{-\xi} e^{-i \psi}\right] A^U_{\psi} \ ,
\\
A^U_{\rho} & =\frac{\partial \psi}{\partial \rho} A^U_{\psi} 
=-\frac{i}{4}\left[e^{-\xi} e^{i \psi}+e^{\xi} e^{-i \psi}-e^{\xi} e^{i \psi}-e^{-\xi} e^{-i \psi}\right] A^U_{\psi} \ .
\end{align}
It follows that at large $\xi$, they behave as
\begin{equation}
 A^U_{\tau} \approx \frac{1}{2} e^{\xi} \cos \psi ~\Phi_\infty, 
 \quad
A^U_{\rho} \approx \frac{1}{2} e^{\xi} \sin \psi ~\Phi_\infty \ .
\end{equation}
This indicates that they exponentially diverge at $\xi \rightarrow \infty$.
One can avoid the divergence by the gauge transformation with
\begin{equation*}
g=e^{-i \Phi_\infty \psi}
\end{equation*}
because $A^U_\psi$ is transformed as
\begin{align}
A^U_\psi \to& g A^U_{\psi} g^{-1}+i g \partial_{\psi} g^{-1}=A^U_{\psi}-\Phi_\infty
\notag 
\\
& =\pm2\left\{\frac{ e^{-\xi}}{e^{\xi}-e^{-\xi}}- \frac{\alpha e^{-\alpha \xi}}{e^{\alpha \xi}-e^{-\alpha \xi}}\right\}\left(h_{1}+\sqrt{3} h_{2}\right) 
\notag
\\
& =O\left(e^{-2 \xi}\right) \quad(\xi \rightarrow \infty)
\label{eq:avoid_div}
\end{align}
for $\alpha>1$.
On the other hand, $A^U_{z}$ is transformed to
\begin{equation}
\begin{aligned}
A^U_z\to g A^U_{z} g^{-1} 
& =\frac{i \bar{z}}{1+|z|^{2}}\left(h_{1}+\sqrt{3} h_{2}\right)-\frac{\sqrt{2} f}{1+|z|^{2}} e^{\pm i(1-\alpha) \psi}\left(e_{-1}-e_{-3}\right)
\end{aligned}
\end{equation}
This is single-valued if and only if $\alpha$ is an integer.
Therefore, if $\alpha$ is an integer, the gauge potentials for the solution can be finite and single-valued everywhere in $\mathbb{R}^4$.

\section{Hyperbolic monopole - hyperbolic antimonopole bound state}
\label{sec:HM-AHM}

In this section, we impose the condition \eqref{eq:cond_HM-HAM}, which leads to the ansatz
\begin{equation}
    \varphi_{1}=\varphi_{3} \equiv f, \quad C_{\psi}^{1}=-\sqrt{3} C_{\psi}^{2}\equiv \calC \ .
    \label{eq:ansatz_HM-AHM}
\end{equation}
The topological charge vanishes identically for this ansatz. Consequently, any nontrivial solution cannot be (anti-)self-dual and instead represents a saddle point of the YM action. 

\subsection{Formulation}

We first discuss the algebraic structure of the ansatz.
With this ansatz, the gauge potential takes the form
\begin{equation}
    \begin{split}
        &A^U_\psi=\calC \left(h_1-\frac{1}{\sqrt{3}}h_2\right) \ ,\\
        & A^U_\xi=0 \ , \\
        &A^U_z=\frac{i\bar{z}}{1+|z|^2}\(h_1+\sqrt{3}h_2\)+\dfrac{\sqrt{2}f}{1+|z|^2}\(e_{-1}-e_{-3}\) \ .
    \end{split}
    \label{eq:gauge_pot_Cf}
\end{equation}
The $z$-component takes values in the principal $su(2)$ subalgebra spanned by the generators in Eq.~\eqref{eq:generator_principal_SU(2)}. By contrast, the $\psi$-component is proportional to
$h_1-h_2/\sqrt{3}$, which lies in the orthogonal complement of the principal $su(2)$ subalgebra.

We write down the eom and action for the ansatz. 
By substituting the ansatz into Eqs. \eqref{eq:eom_gen_C1} and \eqref{eq:eom_gen_C2}, one obtains
\begin{equation}
    \partial_{\xi}\left[\sinh ^{2} \xi \partial_{\xi} \calC\right]-6 \calC f^{2} =0\ ,
    \label{eq:eom_gen_calC}
\end{equation}
and Eqs. \eqref{eq:eom_gen_phi1} and \eqref{eq:eom_gen_phi3} reduce to
\begin{equation}
    \partial_{\xi}^{2} f -\calC^{2} f+\frac{f\left(1-f^{2}\right)}{\sinh^2 \xi}  = 0\ .
    \label{eq:eom_gen_f}
\end{equation}
The action is simplified to the form
\begin{align}
S=32 \pi^2 \int  d \xi \left[ \frac{1}{6} \sinh ^{2} \xi\left(\partial_{\xi} \calC\right)^{2} 
+\left(\partial_{\xi} f\right)^{2}+ \calC^{2} f^{2}+\frac{\left(1-f^{2}\right)^{2}}{2\sinh^2 \xi} \right]
    \label{eq:action_HM-AHM_calC-f}
\end{align}
Note that Eqs.~\eqref{eq:eom_gen_calC} and \eqref{eq:eom_gen_f} can be derived as the EL equations of the system \eqref{eq:action_HM-AHM_calC-f}. 
Unfortunately, the lower bound of the action cannot be given only by topological terms. It implies that the equations \eqref{eq:eom_gen_calC} and \eqref{eq:eom_gen_f} cannot be reduced to first-order ODEs using the Bogomol'ny trick. In addition, they are not integrable. Therefore, we should rely on a numerical technique to solve them.

\subsection{Boundary conditions and asymptotic analysis}
\label{subsec:boundary_conditions}

In order for the action to be finite, we impose the boundary conditions
\begin{equation}
    \calC(0)=0, ~~ f(0)=1, \qquad \lim_{\xi\to\infty}\calC(\xi)=\gamma,~~\lim_{\xi\to\infty}f(\xi)=0
    \label{eq:boundary_cond_HM-AHM}
\end{equation}
with a constant $\gamma$. Since the action is invariant under the sign flip of $\calC$, we let $\gamma>0$ without loss of generality.
The equations can be solved for any value of $\gamma$. However, similarly to the constant $\alpha$ in Sec.~\ref{sec:HmM}, the constant $\gamma$ is restricted to an integer in order that the gauge potentials $A_\mu$ are finite and single-valued everywhere in $\reals^4$.

Following the discussion in Subsec.~\ref{subsec:single-valuedness_HmM}, we show that $\gamma$ has to be an integer so that the gauge potentials are finite and single-valued everywhere in $\reals^4$.
In the coordinates $(\tau,\rho)$ defined in Eq.~\eqref{eq:tau_rho}, the corresponding components of the gauge potential are given by
\begin{equation}
    \begin{split}
        &A^U_\tau=\frac{d\psi}{d\tau}A^U_\psi
        =\[1+\cosh\xi\cos\psi  \]A^U_\psi
         \ ,\\
        &A^U_\rho=\frac{d\psi}{d\rho}A^U_\psi
        =\sinh\xi\sin\psi A^U_\psi \ .\\
    \end{split}
\end{equation}
This indicates that at large $\xi$, they behave as
\begin{equation}
    \begin{split}
        &A^U_\tau\approx\frac{1}{2}e^\xi \cos\psi\Phi_\infty \ , \qquad
        A^U_\rho\approx\frac{1}{2}e^\xi \sin\psi\Phi_\infty \ ,
    \end{split}
\end{equation}
where in this case, $\Phi_\infty$ is defined by
\begin{equation}
    \Phi_\infty=\lim_{\xi\to\infty}A^U_\psi=\gamma\(h_1-\frac{1}{\sqrt{3}}h_2\) .
\end{equation}
Therefore, they clearly diverge at $\xi\to\infty$.
This divergence can be removed by the gauge transformation with 
$g=e^{-i\Phi_\infty\psi}$ similarly to the HmM case~\eqref{eq:avoid_div}, and  $A^U_z$ is transformed as
\begin{equation}
    A^U_z\to gA^U_zg^{-1}
    =\frac{i\bar{z}}{1+|z|^2}(h_1+\sqrt{3}h_2)+\frac{\sqrt{2}f}{1+|z|^2}\(e^{-i\gamma \psi}e_{-1}-e^{i\gamma\psi}e_{-3} \)
    \label{eq:gAzg}
\end{equation}
Eq.~\eqref{eq:gAzg} is single-valued if and only if $\gamma$ is an integer. 
Consequently, $\gamma$ should be an integer in order that the gauge potentials are finite and single-valued everywhere in $\mathbb{R}^4$.

\subsubsection{Asymptotic solution}
\label{subsubsec:asymptotic_sol}

We examine the asymptotic properties of $\calC$ and $f$. At large $\xi$, Eq.~\eqref{eq:eom_gen_f} can be linearized as
\begin{equation}
    \pd_\xi^2f=\gamma^2f \ .
\end{equation}
It follows that $f\approx Ke^{-\gamma\xi}~~(\xi\to\infty)$ with a constant $K$. 
We assume $\calC\approx -L\coth\xi+(\gamma+L)\coth(s\xi)$ with constants $L$ and $s$. Then, substituting the asymptotic form into Eq.~\eqref{eq:eom_gen_calC}, we get
\begin{equation}
    2s(s-1)(\gamma+L)e^{2(1-s)\xi}=6K^2\gamma e^{-2\gamma\xi} \ .
\end{equation}
Thus, we find
\begin{equation}
    s=1+\gamma, \qquad
    K^2=\frac{1}{3}(1+\gamma)(\gamma+L) \ .
\end{equation}
On the other hand, at small $\xi$, we have series solution of the form
\begin{align}
& {\cal C} = c_1 \xi^2 + {\cal O}(\xi^4), \\
& {f} = 1 + c_2 \xi^2 + {\cal O}(\xi^4),
\end{align}
where $c_1$ and $c_2$ are constants, and all higher order coefficients are determined by $c_1$ and $c_2$.

\subsection{\texorpdfstring{Magnetic weight and interpretation as an HM-$\overline{\text {HM}}$ bound state}{Magnetic weight and interpretation as an HM--anti-HM bound state}}

For the ansatz \eqref{eq:gauge_pot_Cf} and the boundary conditions
\eqref{eq:boundary_cond_HM-AHM}, the asymptotic Higgs field and the magnetic
charge matrix are
\begin{align}
    \Phi_\infty &= \frac{\gamma}{3}\diag(1,-2,1) \ , \\
    Q_{\rm M} &= \diag(2,0,-2) \ .
\end{align}
For $\gamma>0$, ordering the eigenvalues of the Higgs field requires the
simultaneous interchange of the second and third entries, which gives
\begin{equation}
    \widetilde{\Phi}_\infty
    =\frac{\gamma}{3}\diag(1,1,-2),
    \qquad
    \widetilde{Q}_{\rm M}=\diag(2,-2,0)=2H_1 \ .
\end{equation}
Therefore, the mass parameters of constituent monopoles are
\begin{equation}
    (m_1,m_2,m_3)=(0,\gamma,\gamma) \ ,
\end{equation}
and the magnetic weights are
\begin{equation}
    (k_1,k_2) = (2,0) \ .
\end{equation}
Note that the sign of the magnetic weights are not unique, because the first two eigenvalues of $\widetilde{\Phi}_\infty$ are degenerate and one can exchange the first and second component of the magnetic matrix.
Under this equivalence relation, such configurations are denoted by
$(k_1,k_2)=([2],0)$.

Because of the equivalence relation, the magnetic weight does not have a unique interpretation as the number of  constituent monopoles. 
We change the basis of the coroot and write
\begin{equation}
    \widetilde{Q}_{\rm M} = -2H_2+2H_3 \ .
\end{equation}
In this basis, we can interpret the configurations as being composed of two type-2 $\AHM$s and two type-3 HMs.
Therefore, the configurations can be interpreted as a bound state of HM and $\AHM$.

For $\gamma<0$, on the other hand, we obtain
\begin{equation}
    \begin{split}
    &\widetilde{\Phi}_\infty
    =\frac{|\gamma|}{3}\diag(2,-1,-1) \ ,
    \\
    &\widetilde{Q}_{\rm M}=\diag(0,2,-2)=2H_2=-2H_1+2H_3 \ .
    \end{split}
\end{equation}
This indicates that the mass parameters and magnetic weight are 
\begin{equation}
    (m_1,m_2,m_3)=(|\gamma|,0,|\gamma|) \ ,
    \\
    (k_1,k_2) = (0,[2]) \ .
\end{equation}
Moreover, each configuration can be viewed as a composite of two type-1
$\AHM$s and two type-3 HMs.

\subsection{Numerical analysis}

%%%%%%%%%%%%%%%%%%%%%%%%%%%%%%%%%%%%%%%%%%
\begin{figure}[t]
    \centering
    \includegraphics[width=1.0\linewidth]{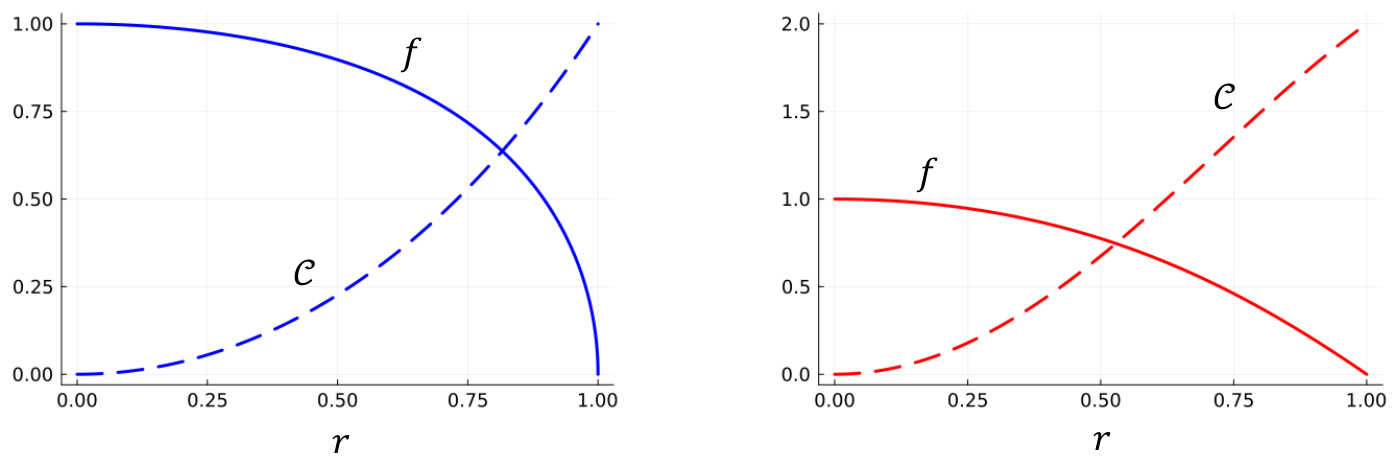}
    \caption{Radial profiles of the numerical solutions 
    as functions of $r=\tanh\xi$.
    Solid and dashed curves represent $f$ and $\calC$ respectively.
    The left panel shows the solutions with $\gamma = 1$, and the right panel shows those with $\gamma = 2$.}
    \label{fig:HM-AHM_profile}
\end{figure}

\begin{figure}[t]
    \centering
    \includegraphics[width=0.5\linewidth]{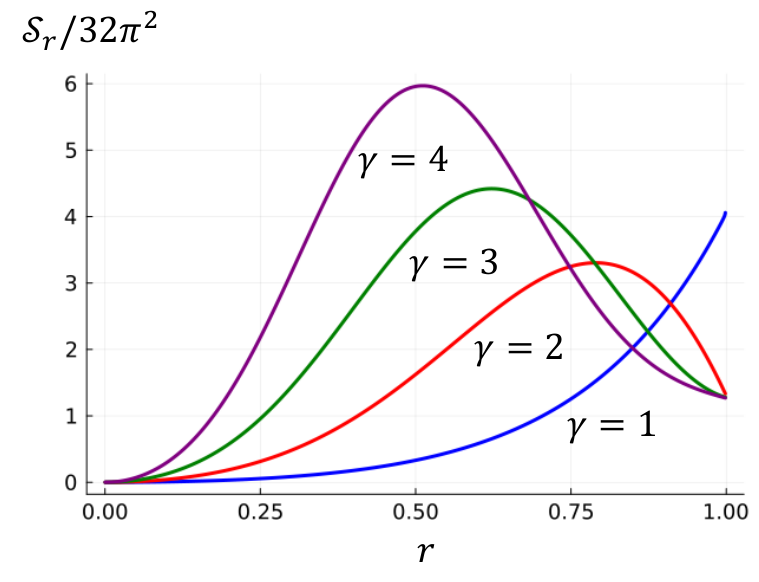}
    \caption{Normalized radial action density ${\cal S}_r/32\pi^2$ of the solutions with $\gamma=1,2,3,4$, where $S=\int dr {\cal S}_r$.
    For $\gamma=1$, the action density attains its maximum at $r=1$, corresponding to the limit $\xi\to\infty$. As $\gamma$ increases, the peak shifts toward smaller values of $r$. }
    \label{fig:HM-AHM_action}
\end{figure}

\begin{figure}[t]
    \centering
    \includegraphics[width=0.5\linewidth]{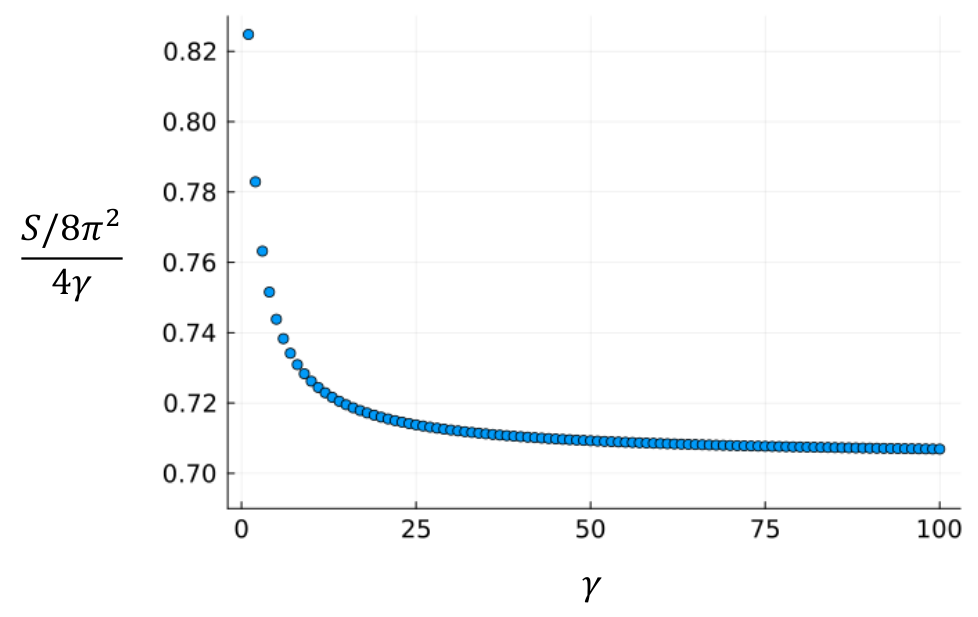}
    \caption{Normalized action $S/32\pi^2\gamma$ as a function of the
    integer parameter $\gamma$. The normalization $32\pi^2\gamma$ is the
    total action of four noninteracting HMs and $\AHM$s. The numerical values decrease monotonically.  
    For $\gamma=100$, $S/32\pi^2\gamma=0.707$.}
    \label{fig:HM-AHM_action-gamma}
\end{figure}
%%%%%%%%%%%%%%%%%%%%%%%%%%%%%%%%%%%%%%%%%%

To conduct a numerical analysis, we use the normalized coordinate \eqref{eq:normalized_coordinate}.
In terms of the coordinate, Eqs.~\eqref{eq:eom_gen_f} and \eqref{eq:eom_gen_calC} can respectively be written as 
\begin{equation}
    \begin{aligned}
& \frac{d^2 \calC}{dr^2}+\frac{2}{r} \frac{d\calC}{dr}-\frac{6\calC f^{2}}{r^{2}\left(1-r^{2}\right)} =0 \ , 
\\
& \frac{d^2f}{dr^2}-\frac{2 r}{1-r^{2}} \frac{df}{dr}-\frac{\calC^{2} f}{\left(1-r^{2}\right)^{2}}+\frac{f\left(1-f^{2}\right)}{r^{2}\left(1-r^{2}\right)}=0 \ .
\end{aligned}
\label{eq:eom_gen_normalized_coord}
\end{equation}
In addition, the action is defined by $S=\int dr {\cal S}_r $ with the density
\begin{align}
\frac{{\cal S}_r}{8\pi^2}=   \frac{r^{2}}{6} \left(\frac{d\calC}{dr}\right)^{2}+\left(1-r^{2}\right)\left(\frac{df}{dr}\right)^{2}
+\frac{ \calC^{2} f^{2}}{1-r^{2}}+\frac{\left(1-f^{2}\right)^{2}}{2r^{2}} \ .
\end{align}

We solved the coupled ODE \eqref{eq:eom_gen_normalized_coord} using the Newton-Raphson method with the second order finite difference approximation, where the system is discretized on a grid with 2001 points. 
In Fig.~\ref{fig:HM-AHM_profile}, we show the profiles of numerical solutions with $\gamma=1,2$.
The action densities of the solutions are represented in Fig.~\ref{fig:HM-AHM_action}. The value of the action density at $r=1$ gradually decreases as $\gamma$ increases, unlike the HmM case. This implies that the constant $L$ in the asymptotic form of $\calC$ depends on $\gamma$. When we compare the HM-$\AHM$ solutions $\gamma=n$ and HmM solutions with $\alpha=n+1$, the shapes of the action densities are very similar, but in the HM-$\AHM$ case the maximum of the action density exists at a larger value of $r$ than in the HmM case. It implies that the size of a HM-$\AHM$ state is smaller than the corresponding HmM configuration.
Fig.~\ref{fig:HM-AHM_action-gamma} shows the $\gamma$ dependence of the total action.  
We normalized the action by that of the single HM or $\overline{\text{HM}}$, i.e., $8\pi^2$, and divided it by $4\gamma$.
The ratio $S/32\pi^2\gamma$ approaches approximately $0.7$ as $\gamma$ increases.  
To elucidate this behavior, we examine the large-$\gamma$ limit.

\subsection{\texorpdfstring{Large-$\gamma$ limit}{Large-gamma limit}}
Let us divide the action into the core and tail contributions,
\begin{equation}
    \hatS=\hatS_{\mathrm{inner}}+\hatS_{\mathrm{outer}} \ ,
\end{equation}
where $\hatS=S/32\pi^2\gamma$. 
According to the asymptotic behaviour of $f$, one can see that the monopole core size is ${\cal O}(\gamma^{-1})$.
Therefore, we define
\begin{align}
    \hatS_{\text{inner}}=\frac{1}{32\pi^2\gamma}\int_0^{\Lambda/\gamma} d\xi~{\cal S},\qquad
    \hatS_{\text{outer}}=\frac{1}{32\pi^2\gamma}\int_{\Lambda/\gamma}^\infty d\xi ~ {\cal S} 
\end{align}
with a parameter $\Lambda$ satisfying  $1\ll \Lambda \ll \gamma$. 

First we analyze the outer part. 
In the outer region, we assume that $f$ has decayed to its vacuum value, $f=0$, and so the action is approximated as
\begin{equation}
    \hatS_{\text{outer}}\approx\gamma^{-1}\int_{\Lambda/\gamma}^\infty d\xi 
    \left[ \frac{1}{6}\sinh^2\xi (\partial_\xi \calC)^2 +\frac{1}{2\sinh^2\xi} \right] \ .
    \label{eq:action_outer_def}
\end{equation}
In addition, following from the asymptotic analysis done in Sec.~\ref{subsubsec:asymptotic_sol} we assume that 
\begin{equation}
    \calC_\text{outer} = \gamma - L(\coth\xi -1) \ ,
    \label{eq:C_outer}
\end{equation}
where $L$ is a constant.
Substituting Eq.~\eqref{eq:C_outer} into Eq.~\eqref{eq:action_outer_def}, one obtains
\begin{equation}
    \hatS_{\text{outer}}
    \approx\frac{\kappa}{\gamma}\left( \coth\frac{\Lambda}{\gamma}-1 \right) = \frac{\kappa}{\Lambda} + {\cal O}(\gamma^{-1})
    \label{eq:action_outer_result}
\end{equation}
where $\kappa=L^2/6+1/2$.

For the inner region, using the approximation $\sinh\xi \approx \xi$, one may write
\begin{align}
    \hatS_{\text{inner}} &\approx\gamma^{-1}\int_0^{\Lambda/\gamma} d\xi~ 
    \left[\frac{1}{6} \xi^{2} \left(\partial_{\xi} \calC\right)^{2} 
+\left(\partial_{\xi} f\right)^{2}+ \calC^{2} f^{2}+\frac{\left(1-f^{2}\right)^{2}}{2\xi^2} \right] 
\notag
\\
&=\int_0^{\Lambda} d\xi'~ 
    \left[\frac{1}{6} {\xi'}^{2} \left(\partial_{\xi'} \calC'\right)^{2} 
+\left(\partial_{\xi'} f\right)^{2}+ {\calC'}^{2} f^{2}+\frac{\left(1-f^{2}\right)^{2}}{2{\xi'}^2} \right] \ ,
\label{eq:inner_rescaled}
\end{align} 
where we defined $\xi'=\gamma\xi$ and $\calC'=\calC/\gamma$. We further divide Eq.~\eqref{eq:inner_rescaled} into two parts:
\begin{equation} 
    \hatS_\text{inner} \approx \hatS_\text{flat}^\text{full}-\hatS_\text{flat}^\text{tail}  
\end{equation}
where
\begin{align}
    &\hatS_\text{flat}^\text{full}=\int_0^{\infty} d\xi'~ 
    \hat{\cal S}_\text{flat} , \qquad
\hatS_\text{flat}^\text{tail}  =\int_\Lambda^{\infty} d\xi'~ \hat{\cal S}_\text{flat} 
\end{align}
with
\begin{equation}
    \hat{\cal S}_\text{flat} = \frac{1}{6} {\xi'}^{2} \left(\partial_{\xi'} \calC'\right)^{2} 
+\left(\partial_{\xi'} f\right)^{2}+ {\calC'}^{2} f^{2}+\frac{\left(1-f^{2}\right)^{2}}{2{\xi'}^2} 
\label{eq:flat_limit}
\end{equation}
Note that Eq.~\eqref{eq:flat_limit} corresponds to the action density in the flat space limit.
The equations of motion associated with the flat limit action are
\begin{align}
    &\partial_{\xi'}\left( {\xi'}^2 \partial_{\xi'} \calC' \right) - 6\calC' f^2 =0 \ ,
    \label{eq:eom_flat_C}
    \\
    &\partial_{\xi'}^2f-{\calC'}^2f + \frac{f(1-f^2)}{{\xi'}^2}=0
    \label{eq:eom_flat_f}
\end{align}
In $\xi'\gg 1$, Eq.~\eqref{eq:eom_flat_C} is linearized as
\begin{align}
    \partial_{\xi'}\left( {\xi'}^2 \partial_{\xi'} \calC' \right) = 0
\end{align}
because $f\approx 0$. Therefore, the asymptotic solution of $\calC'$ is
\begin{equation}
    \calC' \approx 1-\frac{\ell}{\xi'} 
    \label{eq:Ctail_flat}
\end{equation}
The matching condition for Eq.~\eqref{eq:C_outer}
and \eqref{eq:Ctail_flat} at $\xi'=\Lambda$ is 
\begin{equation}
    \ell = L \ .
\end{equation}
Then, using \eqref{eq:Ctail_flat}, one obtains
\begin{align}
    \hatS_\text{flat}^\text{tail} \approx \int_\Lambda^\infty d\xi' 
    \left[
        \frac{1}{6} {\xi'}^{2} \left(\partial_{\xi'} \calC'\right)^{2} 
    +\frac{1}{2{\xi'}^2} 
        \right]
        =\frac{\kappa}{\Lambda} \ .
\end{align}

Combining the above results, we obtain
\begin{equation}
    \hatS \approx \hatS_\text{flat}^\text{full}-\hatS_\text{flat}^\text{tail} + \hatS_\text{outer} = \hatS_\text{flat}^\text{full} + {\cal O}(\gamma^{-1}) \ .
\end{equation}
Therefore, in the large-$\gamma$ limit, the normalized action can be approximated by the energy of a non-Bogomolny monopole on $\reals^3$. Such a solution was constructed in Ref.~\cite{Ioannidou:1999xq}, and its normalized energy is estimated to be $4.3/6\approx0.717$ in our conventions. This value is consistent with our numerical results.

\section{Summary}
\label{sec:summary}

In this paper, we have constructed two types of $S^1$-invariant solutions of
$SU(3)$ YM theory on $\mathbb{R}^4$, which are generalizations of HM. The first describes a
noninteracting cluster of HMs, whereas the second describes a HM-$\AHM$ bound state. 
To obtain these solutions, we formulated an ansatz using the CFN
decomposition and harmonic maps of full map type from $S^2$ into the flag manifold
$F_3=SU(3)/U(1)^2$.
Upon imposing $S^1$ invariance on the ansatz, the YM equations reduce
to a coupled system of ODEs for the radial profile functions.

For the HmM case, the gauge potential takes values in the
principal $su(2)$ subalgebra of $su(3)$.  
The reduced equations coincide with
those of the standard $SU(2)$ HM, while the action and the
topological charge are exactly four times larger than those of the standard $SU(2)$ HM.  
This allowed us to obtain the solutions using the BPS trick.  
The solutions have $|Q|=4n$ and $S=32\pi^2n$, where the standard $SU(2)$ HM has $|Q|=n$ and $S=8\pi^2n$.
Their magnetic weights are $\pm(2,2)$, which implies 
that they describe a cluster of HMs with
two types of fundamental constituents.

In the second case, the topological charge vanishes identically.
Therefore, any nontrivial solution in this sector can be neither self-dual nor anti-self-dual and correspond to saddle points of the YM action.  
The solutions can be interpreted as HM-$\AHM$ bound states.
Regular finite-action solutions have been constructed by numerically solving the coupled ODEs.
Their magnetic weights are given by $([2],0)$ or $(0,[2])$.

An important direction for future work is to explore the physical implications of these solutions, which have no counterparts in $SU(2)$ YM theory.
Such an analysis may reveal nonperturbative phenomena intrinsic to the $SU(3)$ theory. 
It would also be worthwhile to determine the negative-mode
spectra of these saddle points and relax the imposed spherical and circle symmetries.

\ack
The author is grateful to Luiz Agostinho Ferreira and Yakov Shnir for their many helpful comments.
The author also gratefully acknowledges the hospitality of the Instituto de Física de São Carlos, Universidade de São Paulo (IFSC/USP).
This work is supported in part by JSPS KAKENHI [Grants  No.~JP23KJ1881, No.~JP26K17156] and the WPI program ``Sustainability with Knotted Chiral Meta Matter (SKCM$^2$)'' at Hiroshima University.

\appendix

\section*{References}
\bibliographystyle{iopart-num}
\bibliography{refs}

\end{document}